\documentclass{article}
\usepackage {graphicx}
\begin{document}
\baselineskip .75cm 

{\bf Comments on "Stronger Uncertainty Relations for All Incompatible Observables"} \\

Recently, Maccone and Pati \cite{pa.1} derived new uncertainty relations which they claim to be stronger than Heisenberg-Robertson or Schrodinger uncertainty relations. Here we comment that their work is a special case of a more general uncertainty relation and we reexamine some of their conclusions. 

Let us consider a state $|\phi> = |\psi_1>  + i \alpha \, |\psi_2>$ \cite{rn.1}, where $|\psi_1> \equiv (A - <A>) |\psi>$,  $|\psi_2> \equiv (B - <B>) |\psi>$, and $\alpha$ is a free real parameter. $A$ and $B$ are incompatible operators and   $|\psi>$ is the state of a system. From Schwarz inequality,we know that $\parallel \phi \parallel ^2$ $\geq$ $|<\psi^{\perp} |\phi>|^2$, where $|\psi^{\perp}>$ is a normalized arbitrary state orthogonal to  $|\psi>$. It immediately leads to the relation, 
\begin{equation}
  \left[ \Delta A^2 - |<\psi^{\perp} |\psi_1>|^2 \right]  + \alpha^2 \left[ \Delta B^2 - |<\psi^{\perp} |\psi_2>|^2\right] + i \alpha \left[\{  <\psi_1 |\psi_2> - <\psi^{\perp} |\psi_2>   <\psi_1|\psi^{\perp}>\}   - \{  c.c\}  \right] \geq 0\,\,, 
\end{equation}
where $\{c.c\}$ means complex conjugate of previous terms inside the square bracket. Above equation reduces to Eq. (3) of Ref. \cite{pa.1} for $\alpha = \pm 1$. Here, the unknown $\alpha$ may be fixed by minimizing above expression and it reduces to 
\begin{equation}
  \left[ \Delta A^2 - |<\psi^{\perp} |\psi_1>|^2\right] \left[ \Delta B^2 - |<\psi^{\perp} |\psi_2>|^2\right] \geq   
  \frac{1}{4} \left|  <[A,B]> - [\{ <\psi^{\perp} |\psi_2>   <\psi_1|\psi^{\perp}>\}  - \{ c.c\} ]   \right|^2 \,\,,
\end{equation} 
which is just a generalization of Heisenberg-Robertson uncertainty relation. Still, there is an arbitrariness in defining $|\phi>$. We may take  $|\phi> = |\psi_1>  + (\beta + i \alpha) \, |\psi_2>$, with two unknowns and similar procedure leads to generalized Heisenberg-Robertson-Schrodinger uncertainty relations, 
\[
 \left[ \Delta A^2 - |<\psi^{\perp} |\psi_1>|^2\right] \left[ \Delta B^2 - |<\psi^{\perp} |\psi_2>|^2\right] \geq    \frac{1}{4} \left|  <[A,B]> - [\{ <\psi^{\perp} |\psi_2>   <\psi_1|\psi^{\perp}>\}  - \{ c.c\} ]   \right|^2
 \]
\begin{equation}
  + \frac{1}{4} \left|  <\{A,B\}> - 2 <A> <B> - [\{ <\psi^{\perp} |\psi_2>   <\psi_1|\psi^{\perp}>\}  + \{ c.c\} ] \right|^2 \,\,\, . \label{eq:gh}
\end{equation} 
It may be convenient to write it as,
\[
 \Delta A^2  + \Delta B^2 \geq |<\psi^{\perp} |\psi_1>|^2 + |<\psi^{\perp} |\psi_2>|^2 + \mbox{\huge $[$}
 \left|  <[A,B]> - [\{ <\psi^{\perp} |\psi_2>   <\psi_1|\psi^{\perp}>\}  - \{ c.c\} ]   \right|^2 
 \]
\begin{equation}
+ \left|  <\{A,B\}> - 2 <A> <B> - [\{ <\psi^{\perp} |\psi_2>   <\psi_1|\psi^{\perp}>\}  + \{ c.c\} ] \right|^2 \,\, \mbox{\huge $]$} ^{1/2} 
 \,\,\, .  \label{eq:f} 
\end{equation} 

As an example, we may take $|\psi^{\perp}> = \frac{|\psi_1>}{\Delta A}$ or $|\psi^{\perp}> = \frac{|\psi_2>}{\Delta B}$, but it reduces to Heisenberg-Robertson-Schrodinger uncertainty relations \cite{ba.1}. We may also verify using the author's example of spin-1 particle state, $|\psi> = \cos \theta |+> + \sin \theta |->$ and $|\psi^{\perp}>$ may be taken as $|0>$ and we see that both sides of Eq. (\ref{eq:f}) are nonzero and are equal, but both sides of Eq. (\ref{eq:gh}) are zero for any values of $\theta$.  

Our second comment is that for the choice of  $|\psi^{\perp}> = \frac{|\psi_1>}{\Delta A}$ or $|\psi^{\perp}> = \frac{|\psi_2>}{\Delta B}$, Eq. (3) of Ref. \cite{pa.1} also reduces to Heisenberg-Robertson uncertainty relation and hence the lower bound of the inequality is again null or trivial. For the same choice, Eq. (4)  of Ref. \cite{pa.1} becomes meaningless.  

Our conclusion is that author's work is a special case of our general formulation for the values $\alpha = \pm 1$ and $\beta = 0$. We also corrected some of their conclusions.

\noindent 
{\bf Vishnu M. Bannur}  \\
{\it Department of Physics}, \\  
{\it University of Calicut, Kerala-673 635, India.}   
\date{}
                                                                                

\end{document}